\documentclass{article}

\usepackage{lineno}
\usepackage{hyperref}

\usepackage{epsfig}

\usepackage{amsmath}
\usepackage{amstext}
\usepackage{cmap}
\usepackage{mathrsfs}
\usepackage{amssymb}
\usepackage{amsfonts}
\usepackage{algorithm}

\usepackage{graphicx}
\usepackage{caption}
\usepackage{subcaption}

\usepackage{url}

\newtheorem{statement}{Statement}
\newtheorem{definition}{Definition}
\newtheorem{proof}{Proof}

\urldef{\mailsa}\path|nikita.kotlyarov@bk.ru, p.prikhodko@skoltech.com|

\newcommand{\XX}{\mathscr{X}}
\newcommand{\DD}{\mathscr{H}}

\newcommand{\ub}{\mathbf{u}}
\newcommand{\vb}{\mathbf{v}}

\newcommand{\R}{\mathbb{R}}

\newcommand{\E}{\mathbb{E}}
\newcommand{\V}{\mathbb{V}}

\begin{document}

\title{Calibration of Sobol indices estimates in case of noisy output}

\author{
  Nikita Kotlyarov\\
  Institute for Information Transmission Problems\\
  Moscow\\
  \texttt{nikita.kotlyarov@bk.ru} \\
  \and \\
  Pavel Prikhodko \\
  Skoltech\\
  Moscow \\
  \texttt{p.prikhodkop@skoltech.com} \\
}

\maketitle              
\begin{abstract}

This paper presents a simple noise correction method for Sobol' indices estimation. Sobol' indices, especially total Sobol' indices are quite sensitive to the noise in the output and tend to be severly biased (overestimated) if no noise correction is done, which may make their computation meaningless in case of even quite moderate noise levels. Proposed method allows to get approximately unbiased noise free estimation of Sobol' indices at the cost of variance of estimate increase if noise can be represented as a combination of additive and multiplicative stationary noise.
Proposed method is more straightforward than schemes found in the literature and does not introduce any assumptions on the function and noise distribution (except that it assumes noise to be stationary and be a combination of additive and multiplicative). One of the appealing features is that there is actual analytical noise correction expression derived.


\end{abstract}

\section{Introduction}

Computational models became really important in lots of different areas (\cite{Beven2000}, \cite{Dayan2001}, \cite{Burn1}). Often they are so complex that special tools are needed to analyse their behavior.
Sensitivity analysis is the study of how the variations in the output of a mathematical model or system can be apportioned to different variations in its inputs.

Sensitivity analysis includes a wide range of metrics and techniques. Most notable are the Morris method \cite{Morris1991}, linear regression-based methods \cite{Iooss2015}, variance-based methods \cite{Saltelli2008}. Among others, \emph{Sobol' (sensitivity) indices} are a common metric to evaluate the influence of model parameters \cite{Sobol93}. Sobol' indices describe the portion of the output variance explained by different input parameters and combinations thereof. This method is especially useful for the case of nonlinear computational models \cite{Saltelli2010}.

Different approaches are applied to estimate Sobol indices. In \cite{Saltelli1999} A. Saltelli uses the Fourier amplitude sensitivity test (FAST) algorithm, which requires to generate complex deisgn of experiment, but is considered to be the most accurate approach. In \cite{Plischke2010} E. Plischke proposed an EASI algorithm, that allows to efficiently compute Sobol indices on any given sample. In \cite{Glen2012} G. Glen demonstrates correlation based method with the correction to the spurious correlation, which is very convenient to compute confidence intervals for the indices and for handling Not-a-number-values in the design.

Generally when starting an analysis two types of Sobol indices are computed:

\begin{enumerate}
\item {\bf Main indices} ($S_i$) which take into account only sole influence of each feature $i$ while all others being fixed. These indexes tell what portion of output variance would be described by considered input provided all other inputs are fixed at their mean values.
\item {\bf Total indices} ($T_i$) which take into account interactions between features. These indexes tell what portion of output variance would be lost if we fix considered feature $i$ to its mean value, while still vary others.
\end{enumerate}

It's important that all Sobol' indices estimates are very sensitive to the noise in the output {\it (or some uncontrollable input parameters that change arbitrary on different function calls)}. Having even moderate noise leads to significant overestimation of main and especially total indices making analysis futile, provided no noise correction is done.

When doing the sensitivity analysis expert usually knows if his function is noisy beforehand, so it makes sense to have special estimation procedure in the setting with noisy output.

In \cite{Liu} here is a review of Probabilistic Sensitivity Analysis (PSA) --- study the impact of uncertainties in design variables and noise parameters on the probabilistic characteristics of a design performance. There were some attempts in the literature aimed to introduce noise correction to the Sobol indices estimation. In \cite{Ioss} authors tried to directly estimate influence of noise and focuses on models that depend on scalar parameter vector X and involve some stochastic process simulations or random fields $\varepsilon(u)$ as input parameters, where $u$ can be spatial coordinates, time scale or any other physical parameters. Fort and others (\cite{Fort}) consider $\mathbb{H}$ is a separable Hilbert space endowed with the scalar product $\langle\cdot,\cdot\rangle$ and a linear regression model $Y = \mu + \sum^{p}_{k=1}\langle\beta_k, X_k\rangle+\varepsilon$, where for $X_i$, $i=1,\ldots,p$ --- $p$ independent centered, $\mathbb{H}$-valued, stochastic processes, $\beta_i$, $i=1,\ldots,p$ are elements of $\mathbb{H}$, $\mu\in\mathbb{R}$ and $\varepsilon$ is a centered noise independent of the processes $X_1, \ldots, X_p$. They construct natural estimators of the Sobol indices for whom they prove asymptotic normality and efficiency, using Karhunen-Loeve decomposition of the processes $X_k$. Lewandowski (\cite{Lewandowski}) explored correlation ratio, estimate of sensivity, based on Sobol indices, and applied it to model $G = f(X) + Y $, where $X$ is the explaining variable and $Y$ is added noise. Interesting results on approximation for features of ANOVA-decomposition components are presented in \cite{Wang}. In this work authors consider ANOVA-decomposition (base for Sobol indices theory) and model of function with additive noise. One more work with same theme --- \cite{Jin}. Function model with additive noise is considered in works on sensitivity analysis not for only Sobol indices, for example, see \cite{Roustant}.

In this article we propose simple and straightforward method to remove noise induced distortion from Sobol' indices estimates. The idea of the method is to introduce virtual input for which we know function does not really depend on and variations of the output one obervers when one changing this input are only due to noise.

Such trick allows one to derive analytical noise correction formulas for main and total Sobol' indices. Such approach works in case noise in the data is stationary and is a combination of additive and multiplicativ components. Also we show that if noise has more complex structure no precise noise correction procedure can be done.

The advantages of proposed approach are:
\begin{enumerate}
\item It is very simple and straightforward to apply.
\item It does not depend on estimation approach, so expert can still select the one suitable for the task.
\item It does not introduce additional source of error (unlike meta models, which often bring to the analysis additional distortions).
\end{enumerate}

The paper is organised as follows:

\begin{itemize}
\item We continue with more formal definition of Sobol indices in section \ref{definitions}.
\item After that we introduce proposed noise correction approach in section \ref{noise_correction}.
\item Finally we show some experimental results in section \ref{experiments} and provide summary in section \ref{conclusion}.
\end{itemize}

\section{Sensitivity Indices}
\label{definitions}

\begin{definition}
Let us define a computational model $Y = F(\vec{X})$, where

$\vec{X} = \Big(X_1, \ldots, X_n\Big) \in \XX \subset \R^d$~is a vector of input variables (or parameters or features),

$Y \in \R^1$ is an output variable and $\XX$ is a design space. The model $F(\vec{X})$ describes the behavior of some physical system of interest.

\end{definition}

{\it {\bf Note} that in the definition above one may consider noise as just one of the inputs $x$ whose value is not controlled by experimenter, is unknown and changes arbitrary. So it is possible to introduce noise in the framework without adding complexity to expressions.}\\

Assuming that the function $F(\vec{X})$ is square-integrable with respect to distribution $\DD$ ({\it i.e.} $\E [F^2(\vec{X})] < +\infty)$), we have the following unique Sobol' decomposition of $\vec{Y} = F(\vec{X})$ (see \cite{Sobol93}) given by
\[
F(\vec{X}) = F_0 + \sum_{i=1}^{d} F_i(\vec{X}_i) + \sum_{1\leq i \leq j \leq d} F_{ij}(\vec{X}_i, \vec{X}_j) + \ldots + F_{1 \ldots d}(\vec{X}_1, \ldots, \vec{X}_d),
\]
which satisfies:
\[
\E[F_\ub(\vec{X}_\ub)F_\vb(\vec{X}_\vb)] = 0, \; \text{if} \; \ub \neq \vb,
\]
where $\ub$ and $\vb$ are index sets: $\ub,\vb \subset \{1, 2, \ldots, d\}$.

Due to orthogonality of the summands, we can decompose the variance of the model output:
\[
D = \V[F(\vec{X})] = \sum_{\substack{\ub \subset \{1, \ldots, d\}, \\ \ub \neq \mathbf{0}}  } \V[F_\ub(\vec{X}_\ub)] = \sum_{\substack{\ub \subset \{1, \ldots, d\}, \\ \ub \neq \mathbf{0}}  } D_\ub,
\] 

In this expression $D_\ub \triangleq \V[F_\ub(\vec{X}_\ub)]$ is the contribution of summand $F_\ub(\vec{X}_\ub)$ to the output variance.

\begin{definition}
     {\it The sensitivity index (Sobol' index)} of variable set $\vec{X}_\ub, \; \ub \subset \{1, \ldots, d\}$ is defined as
     \begin{equation*}
    S_{\ub} = \frac{D_\ub}{D}.
\end{equation*}
\end{definition}

The sensitivity index $S_\ub$ describes the amount of the total variance explained by the variance in the subset of model input variables $\vec{X}_\ub$.

\begin{definition}

Main ($S_i$) and total ($T_i$) Sobol' indices are defined as
$$
S_i \triangleq S_{\{i\}}, i=1,\ldots,d.
$$
$$
T_i \triangleq \sum_{\substack{\ub \subset \{1, \ldots, d\}, \\ i \in \ub}  } S_\ub, i=1,\ldots,d.
$$

\end{definition}

\section{Noise correction formula}
\label{noise_correction}

\begin{definition}
\label{noise_model}
Let us consider the following working model of noisy output:

$$
\vec{Y} = F\left(X_1, X_2, \ldots, X_n, t\right) = \left(1+\alpha\right)G\Big(\vec{X}\Big) + \beta,
$$
where $\alpha$ and $\beta$ are some random noise, $t$ is some virtual variable generating noise (i.e. seed of random generator),

\end{definition}

\begin{definition}
Let us denote $\overline{X_i} \triangleq \Big\{ X_1, \ldots, X_{i-1}, X_{i+1}, \ldots, X_n \Big\}$ (i.e. set of all features except $i$-th).
\end{definition}

\begin{definition}

Let $S_i^\varepsilon$ and $T_i^\varepsilon$ be mean and total Sobol' indices computed for function $F$ (i.e. with noise in $\vec{Y}$ present).

\end{definition}

\begin{definition}

Let $S_i$ and $T_i$ be mean and total Sobol' indices computed for function $G$ (i.e. $\vec{Y}$ with noise removed).

\end{definition}

Having defined necessary concepts we can state the noise correction procedure proposed in the paper.\\

\begin{statement}

If dependency has same structure as in Definition \ref{noise_model} and $\alpha$ and $\beta$ do not depend on $\vec{X}$. Then it holds that:

\begin{equation}
\label{eq1}
S_i = \frac{S_i^\varepsilon}{1 - T_t^\varepsilon}
\end{equation}
\begin{equation}
\label{eq2}
T_i = \frac{T_i^\varepsilon - T_t^\varepsilon}{1 - T_t^\varepsilon}
\end{equation}

\end{statement}

\begin{proof}

Let us calculate Sobol indices for function $F$:

$$
S_i^\varepsilon = \frac{\V(\E(F|X_i))}{\V F} = \frac{\V(\E((1+\alpha)G+\beta|X_i))}{\V F} =
$$
$$
= \frac{\V(\E(1+\alpha)\E(G|X_i)+\E\beta)}{\V F} = \frac{(1+\E\alpha)^2\V(\E(G|X_i))}{\V F}
$$

$$
T_i^\varepsilon = 1 - \frac{\V(\E(F|\overline{X_i}))}{\V F} = 1 - \frac{\V(\E((1+\alpha)G+\beta|\overline{X_i}))}{\V F} =
$$
$$
= 1 - \frac{\V(\E(1+\alpha)\E(G|\overline{X_i})+\E\beta)}{\V F}
= 1 - \frac{(1+\E\alpha)^2\V(\E(G|\overline{X_i}))}{\V F}
$$

Besides, we can consider $T_t^\varepsilon$:
$$
T_t^\varepsilon = 1 - \frac{\V(\E(F|X_1, \ldots, X_n))}{\V F} = 1 - \frac{\V(\E((1+\alpha)G+\beta|X_1, \ldots, X_n))}{\V F} =
$$
$$
= 1 - \frac{\V(\E(1+\alpha)\E(G|X_1, \ldots, X_n)+\E(\beta))}{\V F} = 1 - \frac{(1+\E\alpha)^2\V(G)}{\V F}
$$

Then

$$
1 - T_t^\varepsilon = \frac{(1+\E\alpha)^2\V(G)}{\V F}
$$

From that one can easily derive $S_i = \frac{S_i^\varepsilon}{1 - T_t^\varepsilon}$ and
$T_i = \frac{T_i^\varepsilon - T_t^\varepsilon}{1 - T_t^\varepsilon}$. $\blacksquare$

\end{proof}

Note that results presented above can be used with any indices estimation schema and noise distribution. The only requirement is the stationariness of noise.

Also note that to compute noise corrected Sobol' indices one should use several estimates of noised indices. Which would lead no increased variance of final estimate.

\begin{statement}

If the used Sobol' indices estimation procedure $\,\widehat{\mathbf{\cdot}}\,$ for estimation of $S_i^\varepsilon$, $T_i^\varepsilon$, $T_t^\varepsilon$ is unbiased and estimation errors of $\widehat{S_i^\varepsilon}$, $\widehat{T_i^\varepsilon}$, $\widehat{T_t^\varepsilon}$ are uncorrelated, then estimates of $\widehat{S_i}$, $\widehat{T_i}$ from equations \eqref{eq1} and \eqref{eq2} the bias (first order correction) of estimators:

\begin{equation}
bias\left(\widehat{S_i}\right) = S_i\frac{\V\left(\widehat{T_t^\varepsilon}\right)}{(1-T_t^\varepsilon)^2},
\end{equation}

\begin{equation}
bias\left(\widehat{T_i}\right) = T_i\frac{\V\left(\widehat{T_t^\varepsilon}\right)}{(1-T_t^\varepsilon)^2}.
\end{equation}

And corresponding variances are:

\begin{equation}
\label{eq3}
\V\left(\widehat{S_i}\right) = \frac{\V\left( \widehat{S_i^\varepsilon}\right) + S_i\V\left( \widehat{T_t^\varepsilon}\right)}{1-T_t^\varepsilon},
\end{equation}

\begin{equation}
\label{eq4}
\V\left(\widehat{T_i}\right) = \frac{\V\left( \widehat{T_i^\varepsilon}\right) + \left(1 - T_i\right)\V\left( \widehat{T_t^\varepsilon}\right)}{1-T_t^\varepsilon}.
\end{equation}

\end{statement}

\begin{proof}

Estimators in equations \eqref{eq1} and \eqref{eq2} are common ration estimators. Expressions for bias and variance for such estimators can be found here \cite{Scott1981}. $\blacksquare$

\end{proof}

Looking at equations \eqref{eq3} and \eqref{eq4} one may see that due to a denominator the greater noise level is the greater the variance of noise corrected estimator would be. This effect is accordance with experimental results presented in Section \ref{experiments}.

If method from \cite{Glen2012} is used to compute Sobol' indices $\V\left( \widehat{S_i^\varepsilon}\right)$ can be computed via bootstrap and corresponding corrections to the $S_i$, $T_i$ estimates could be made.

\section{Experiments}
\label{experiments}

In this section we demonstrate how our approach performs on number of examples.

We will use method from \cite{Glen2012} with implementation taken from library \cite{GTSDA} (library version 6.3 was used).

\subsection{Example 1: Linear function}

At first let us consider linear function
$$f(x_1, x_2, x_3, x_4) = 3x_1 + 2x_2 + x_3$$

True values of Sobol indices are equal to $\{9.0/14, 4.0/14, 1.0/14, 0.0\}$ for both main and total indices as no interactions are present.

We set noise parameters to be $\alpha\sim U(0,1), \beta\sim U(0,3)$.

\begin{figure}[H]
\centering
\includegraphics[width=1.1\textwidth]{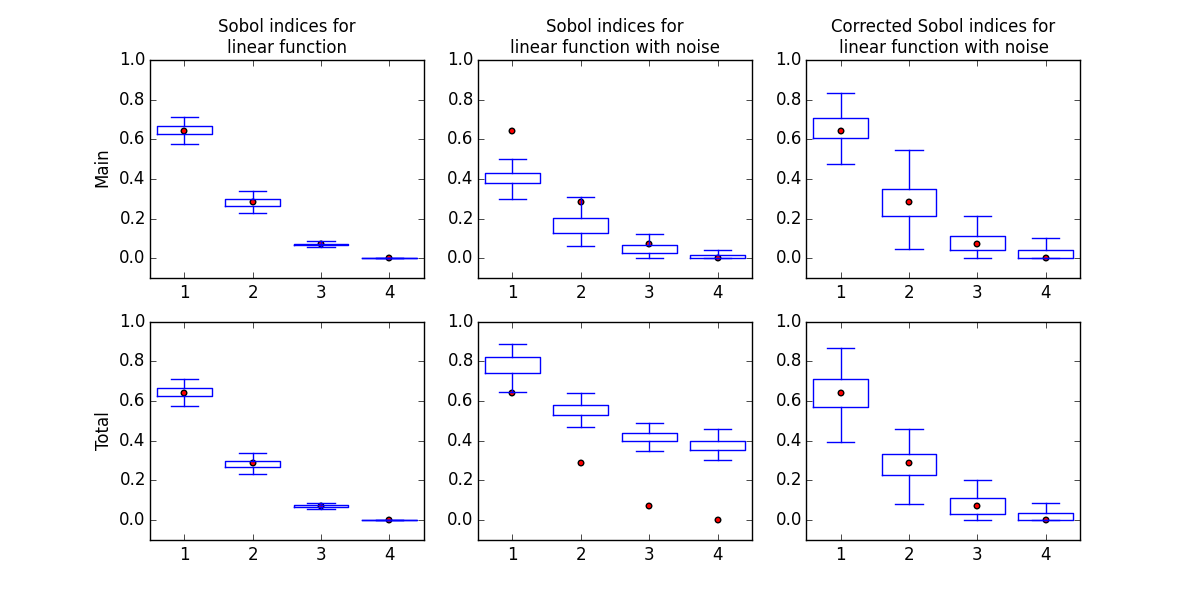}
\caption{Estimates for Sobol indices for linear function and budget 2000. Red dots show true values of indices if noise removed.}
\label{fig:f1}
\end{figure}

\begin{figure}[H]
\centering
\includegraphics[width=1.1\textwidth]{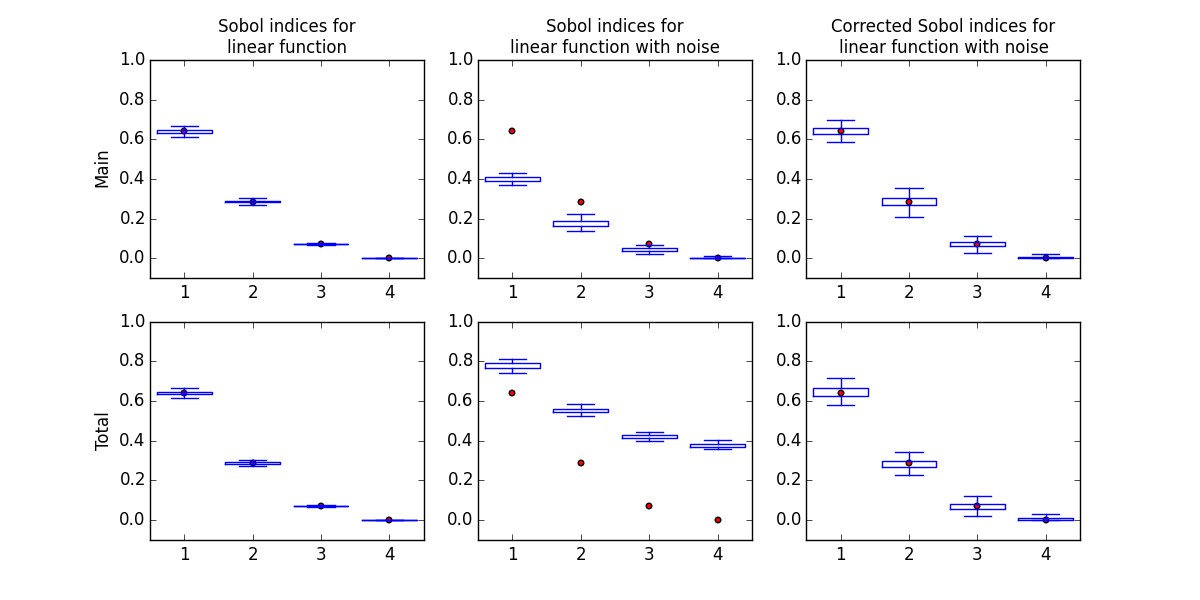}
\caption{Estimates for Sobol indices for linear function and budget 20000. Red dots show true values of indices if noise removed.}
\label{fig:f2}
\end{figure}

Plots showing estimation results for main and total indices are shown on figures~\ref{fig:f1} and~\ref{fig:f2}. Left column shows performance of estimator if no noise were present, middle column shows estimation results for noisy function if no noise correction was done, right column shows results of noise corrected estimator. Red dots show true values of indices if noise removed.

\subsection{Example 2: Sobol' function}

Another common function to consider in papers about Sobol' indices is Sobol' function (it's properties was studied in \cite{Sobol93}).

$$g(x_1,x_2,x_3,x_4,x_5,x_6) = \prod_{i=1}^6\frac{|4x_i-2|+a_i}{1+a_i},$$

where $a=\{0, 0.5, 3, 9, 99, 99\}$.

It's main indices are approximately equal to $\{0.586781$, $0.260792$, $0.0366738$, $0.00586781$, $0.00005868$, $0.00005868\}$ and total indices are $\{0.690086$, $0.356173$, $0.0563335$, $0.00917058$, $0.00009201$, $0.00009201\}$.

We set noise parameters to be $\alpha\sim U(-0.25, 0.25), \beta\sim U(-1, 1)$.

\begin{figure}[H]
\centering
\includegraphics[width=1.1\textwidth]{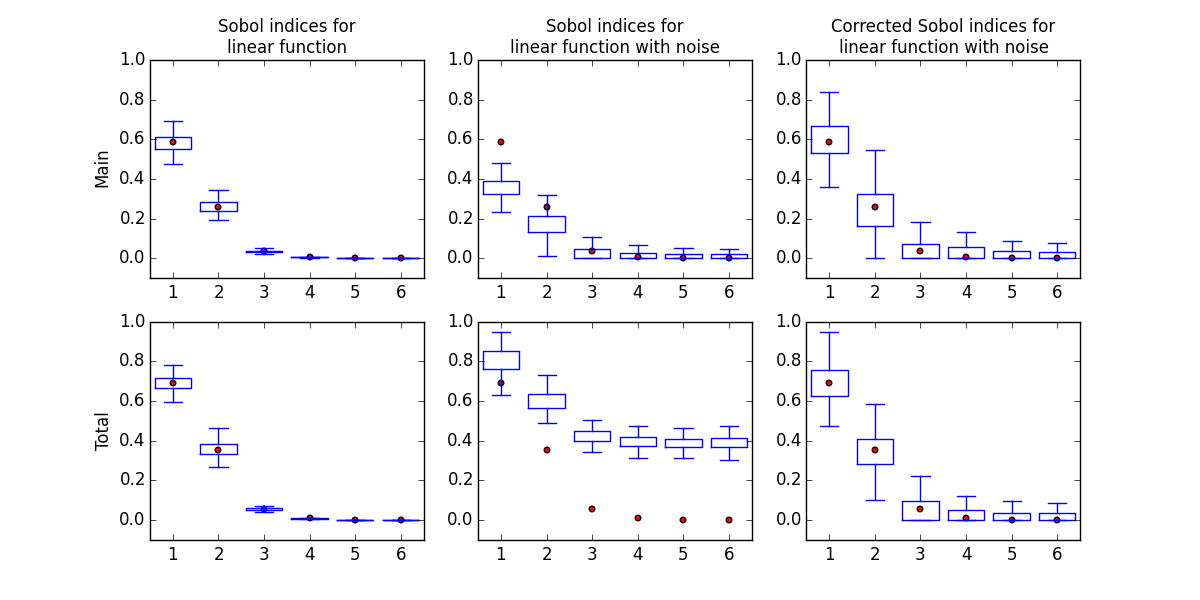}
\caption{Estimates for Sobol indices for Sobol' function and budget 2000. Red dots show true values of indices if noise removed.}
\label{fig:f3}
\end{figure}

\begin{figure}[H]
\centering
\includegraphics[width=1.1\textwidth]{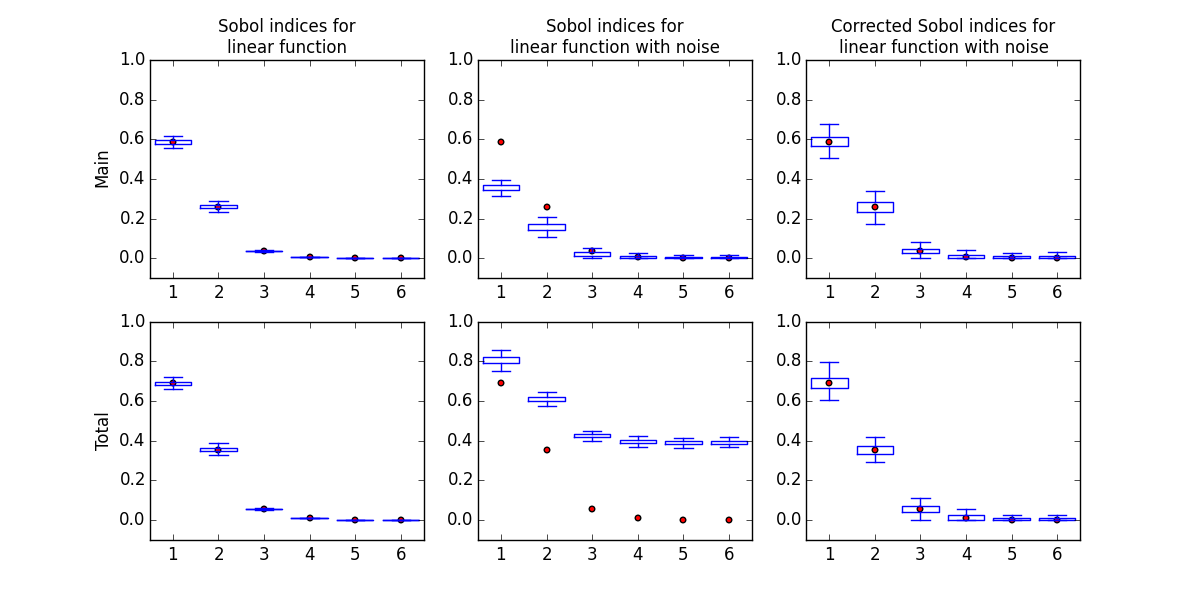}
\caption{Estimates for Sobol indices for Sobol' function and budget 20000. Red dots show true values of indices if noise removed.}
\label{fig:f4}
\end{figure}

Plots showing estimation results for main and total indices are shown on figures~\ref{fig:f3} and~\ref{fig:f4}. Left column shows performance of estimator if no noise were present, middle column shows estimation results for noisy function if no noise correction was done, right column shows results of noise corrected estimator.

For both examples one may see that corrected estimates have much lower bias, but increased variance, which however decreases as sample size gets bigger.

\subsection{Example 3: Steel column under stress.}

Let us consider a bit more complex example -- model a steel column under stress. This function is used to demonstrate uncertainty quantification, for example, in works \cite{Eldred2008}, \cite{Kuschel1997}. The dependency can be written as following formlua:

$$
h(P_1, P_2, P_3, B, D, H, F_0, E, F_s) = F_s - P(\frac{1}{2BD} + \frac{F_0E_b}{BDH(E_b-P)}),
$$
where

$$
P = P_1 + P_2 + P_3,
$$

$$
E_b = \frac{\pi^2EBDH^2}{2L^2}
$$

The steel column stress depends on 9 different parameters, see table \ref{tab1}.



\begin{center}

\begin{tabular}{|c|c|c|c|}

\hline
Parameter & Distribution & Meaning & Measure \\
\hline
$P_1$ & $U(450000, 50000)$ & dead weight load & N \\
\hline
$P_2$ & $U(600000, 100000)$ & variable load & N \\
\hline
$P_3$ & $U(600000, 100000)$ & variable load & N \\
\hline
$B$ & $U(300, 9)$ & flange breadth & mm \\
\hline
$D$ & $U(20, 2)$ & flange thickness & mm \\
\hline
$H$ & $U(300, 15)$ & profile height & mm \\
\hline
$F_0$ & $U(22.5, 7.5)$ & initial deflection & mm \\
\hline
$E$ & $U(210000, 10000)$ & Young's modulus  & MPa \\
\hline
\end{tabular}


\end{center}

In the experiment we introduce noise to the model by assuming that we can not control variable. Firstly, we estimate Sobol indices, using $F_s = 500$ and will consider them as true values in the analysis..

Then we add fictive variable and set random values to $F_s$ on each evaluation (with distribution $U(465, 535)$).

Obtained estimates of Sobol indices are given below. First column show estimates of function without noise, second --- estimates of function with noise (and sobol indices for fictive variable too), and third --- corrected estimates of function with noise.

\begin{figure}[H]
\centering
\includegraphics[width=1.1\textwidth]{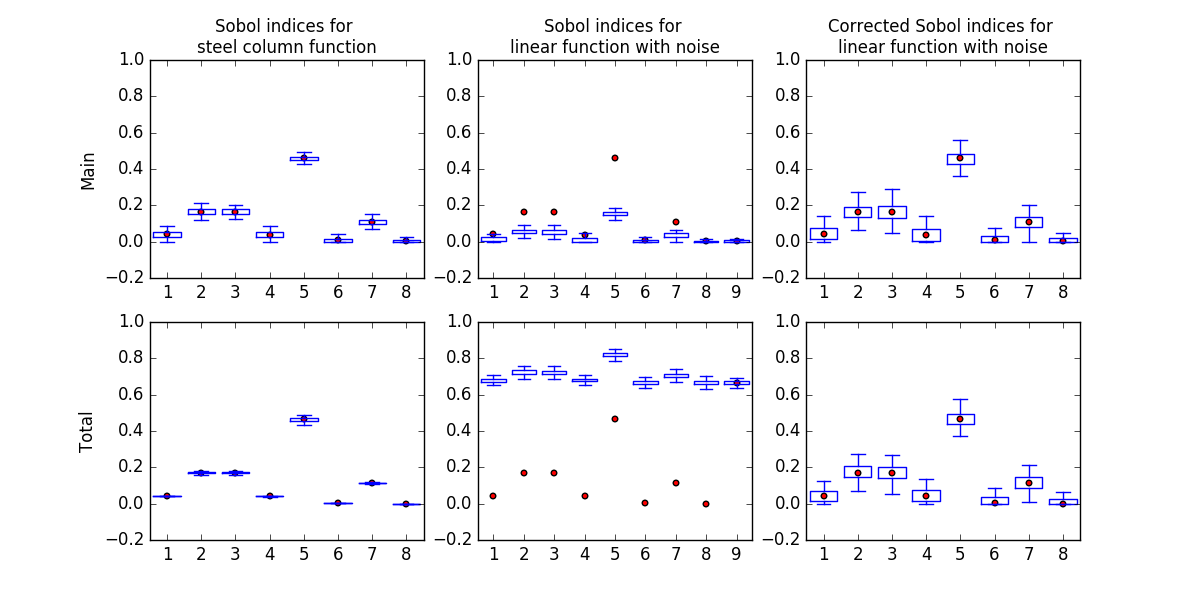}
\caption{Estimates for Sobol' indices for steel column under stress function and budget 50000. Red dots show true values of indices if noise removed.}
\end{figure}

\section{Conclusion}
\label{conclusion}

We presented a noise correction method for Sobol' indices computation and demonstrated it's performance on several toy examples. We think the method is easier to use and, because of its simplicity, more reliable than existing alternatives when one needs to perform sensitivity analysis on the noisy functions (which often happens with numeric solvers where noise may be introduced due to convergence issues).

Method can be applied on top of any Sobol' indices estimatior and usage of explicit formulas to do noise correction allows for straightforward errors propagation in case confidence intervals of original Sobol' indices estimator can be computed.

Main current limitation of the method is its imposed assumption on the noise structure, which in some cases may be not possible to easily check a priory. Though we hope that we will be able to consider more general setting in the future.

One particular possible way to loosen the requirements of noise may be through analysis of higher order indices which method from \cite{Glen2012} allows to do without additional computational budget.

Also it was observed above that noise correction procedure tends to increase variance of corrected indices estimates. Due to that it may make sense to use additional variance reduction techniques in base estimation procedures. One of the promising approaches may be in using of special designs of experiments that reduce the variance of estimation \cite{Burnaev2016}.

\section{Acknowledgements}

The research of the first author was conducted in IITP RAS and supportedsolely  by  the  Russian  Science  Foundation  grant  (project  14-50-00150).


\end{document}